\documentclass[a4paper]{article}
\usepackage{graphicx} 
\usepackage{epstopdf}
\usepackage{textcomp}
\usepackage{amssymb}
\usepackage{INTERSPEECH2022}
\ninept
\title{Spatial-aware Speaker Diarization for Multi-channel Multi-party Meeting}
\name{\begin{tabular}{c}Jie Wang$^1$, Yuji Liu$^{3,4}$, Binling Wang$^2$, Yiming Zhi$^2$, Song Li$^1$, Shipeng Xia$^2$,\\ 
 Jiayang Zhang$^{3,4}$, Feng Tong$^{3,4}$, Lin Li$^{1*}$, Qingyang Hong$^{2*}$\thanks{*Corresponding authors.}\end{tabular}}
\address{
  $^1$School of Electronic Science and Engineering, Xiamen University, China\\
  $^2$School of Informatics, Xiamen University, China\\
  $^3$College of Ocean and Earth Sciences, Xiamen University\\
  $^4$Key Laboratory of Underwater Acoustic Communication and \\
  {Marine Information Technology (Xiamen University), Minister of  Education}}
\email{\{lilin,qyhong\}@xmu.edu.cn}

\begin{document}

\maketitle
\begin{abstract}
This paper describes a spatial-aware speaker diarization system for the multi-channel multi-party meeting. The diarization system obtains direction information of speaker by microphone array. Speaker-spatial embedding is generated by x-vector and s-vector derived from superdirective beamforming (SDB) which makes the embedding more robust. Specifically, we propose a novel multi-channel sequence-to-sequence neural network architecture named discriminative multi-stream neural network (DMSNet) which consists of attention superdirective beamforming (ASDB) block and Conformer encoder. The proposed ASDB is a self-adapted channel-wise block that extracts the latent spatial features of array audios by modeling interdependencies between channels. We explore DMSNet to address overlapped speech problem on multi-channel audio and achieve 93.53\% accuracy on evaluation set. By performing DMSNet based overlapped speech detection (OSD) module, the diarization error rate (DER) of cluster-based diarization system decrease significantly from 13.45\% to 7.64\%.
\end{abstract}


\section{Introduction}

As the applications of speech signal processing becoming more and more popular, the technologies, such as automatic speech recognition (ASR), speaker diarization are facing many challenges in real-world scenarios. In particular, meeting scenario is almost the most challenging and valuable because of its complexity and diversity, including overlapped speech, robustness of speaker embedding, reverberation and unknown number of speakers, etc. 

Speaker diarization has traditionally been addressed as a single-channel problem, in which multi-channels of audios are transformed into mono signal by speech enhancement techniques such as beamforming methods \cite{anguera2007acoustic}\cite{raj2021integration}\cite{boeddeker2018front}. Multi-channel audio samples are collected by microphone array that has abundant spatial location information which is beneficial to OSD task \cite{chen2021overlapped} and speaker representation \cite{kang2020multimodal}. After performing beamfoming, the information will decay or even lost in back-end process of speaker diarization system including OSD and clustering. In this paper, to optimize this problem, we design a spatial-aware multi-channel multi-party speaker diarization system for meeting scenario.

\begin{table}[]
\centering
\caption{Details of AliMeeting training and evaluation data \cite{yu2021m2met}. The audio samples were collected in various room with different size.}
\label{tab:AliMeeting}
\begin{tabular}{lll}
\hline
 & Train & Eval \\ \hline
Duration(h) & 104.75 & 4.00 \\
Session & 212 & 8 \\
Room & 12 & 5 \\
Participant & 456 & 25 \\
Overlap Ratio(Avg.) & 42.27\% & 34.20\% \\ \hline
\end{tabular}
\vspace{-0.4cm}
\end{table}

\begin{figure*}[htb]
  \centering
  \centerline{\includegraphics[width=14.8cm]{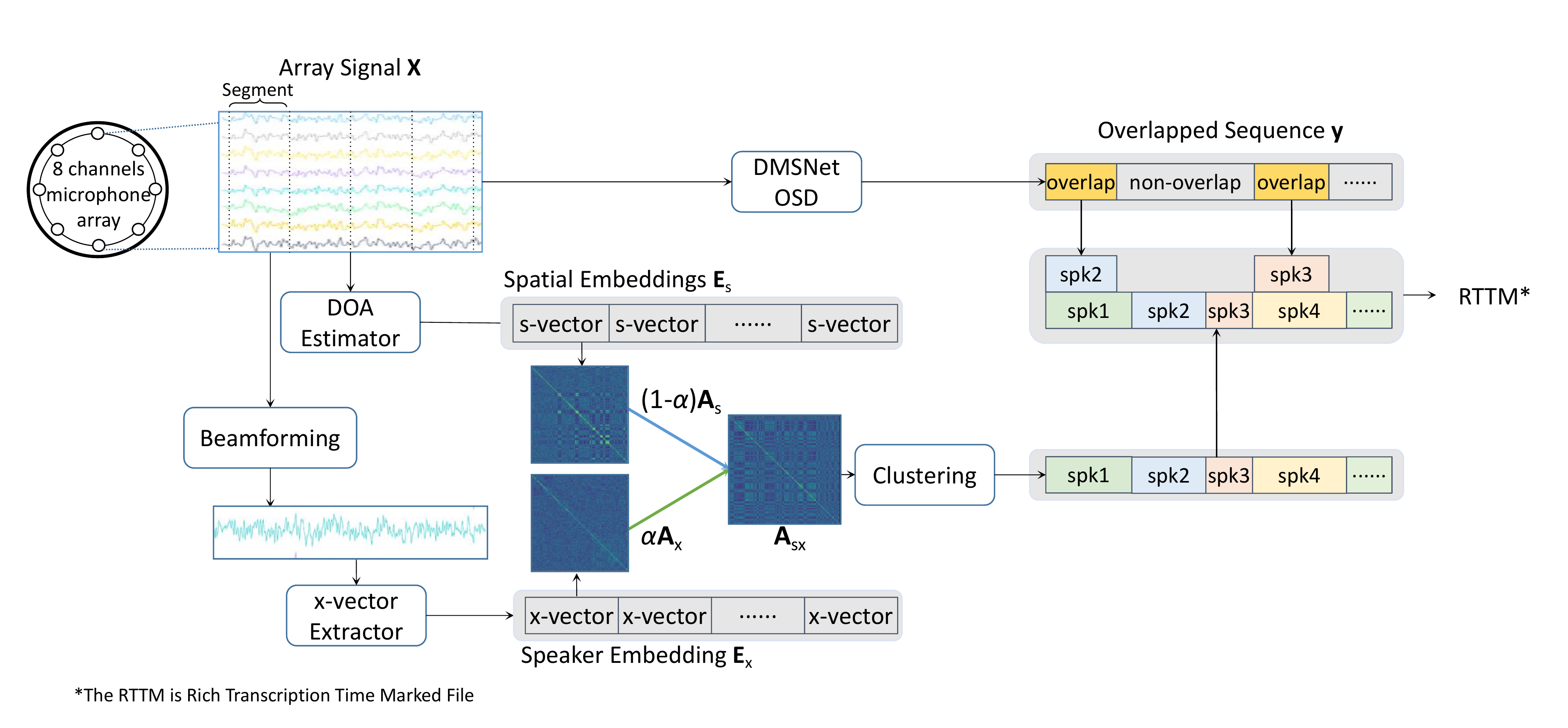}}

\caption{Spatial-aware speaker diarization system overview. The Rich Transcription Time Marked (RTTM) is the output of speaker diarization system.}
\label{fig:system_overview}
\vspace{-0.4cm}
\end{figure*}

We evaluated our diarization system on a sizeable real-recorded Mandarin meeting corpus called AliMeeting which was provided by the Multi-channel Multi-party Meeting Transcription Challenge (M2MeT) \cite{yu2021m2met}. M2MeT challenge focuses on addressing the diarization problem in real-world multi-speaker meetings. 

One of the difficulties in AliMeeting is overlapped speech detection. As shown in Table~\ref{tab:AliMeeting}, The average speech overlap ratio of Train and Eval sets are 42.27\% and 34.76\%, respectively, which make this task more challenging. Overlapped speech not only cause false prediction of speakers number when clustering, but also increases the DER of system directly. In \cite{geiger2013detecting}, a long short-term memory (LSTM) neural network has been applied to OSD. Furthermore, in order to detect the overlapped speech of speakers, many investigations have been carried out \cite{andrei2019overlapped}\cite{cornell2020detecting}\cite{zhang2021investigation}. H Bredin et al. \cite{bredin2020pyannote} applied a Bi-LSTM based OSD named PyanNet to address single-channel overlapped speech. However, the audios of AliMeeting is 8-channel far-field recording with rich spatial information, and the performance of Bi-LSTM based OSD still have room to improve. To tackle this challenging problem, we propose DMSNet for overlapped speech detection. DMSNet is a novel multi-channel sequence-to-sequence architecture used for sequence labeling task which consists of an ASDB block that extract spatial feature and a Conformer \cite{gulati2020conformer} encoder. Compared with Bi-LSTM based OSD model, DMSNet reduces Detection Error Rate (DetER) \cite{Herv2017pyannote} from 42.57\% to 32.47\% on AliMeeting evaluation set. What’s more, the audios of AliMeeting are collected from different conference rooms which are full of noise and reverberation. It is worth noting that the speakers of meeting are required to remain in the same position during recording. In order to make full use of spatial information provided by the microphone array, we fused the spatial-vector (s-vector) and x-vector of speakers to make the similarity matrix of speakers more robust. We performed direction-of-arrival (DOA) technology to extract s-vector. After stacking our methods, the DER of diarization system was reduced to 7.64\% on AliMeeting evaluation set.

The rest of this paper is organized as follows. Section~\ref{sec:system_overview} gives the overview of our speaker diarization system. In Section~\ref{sec:data_preparation}, we describe training set and evaluation set of each module. Experimental setup and results are shown in Section~\ref{sec:experimental_result}. Finally, we conclude our work in Section~\ref{sec:conclusions}.

\section{System overview}
\label{sec:system_overview}
As shown in Figure~\ref{fig:system_overview}, our speaker diarization system consists of speaker embedding extractor, spatial embedding extractor (DOA estimator), clustering modules and OSD. We segment audio samples according to the oracle VAD label provided by M2MeT. We will describe each module in detail as follows.

\subsection{Speaker embedding}

The ResNet34-SE \cite{zhou2019deep} is employed as the x-vectors extractor with additive margin softmax loss \cite{wang2018additive} (AAM-softmax) which learns a segment-level representation from the input acoustic feature. The dimension of x-vector is reduced from 256 to 128 by Linear Discriminant Analysis (LDA). In our system, the input is 81-dimensional log-mel filter-bank (FBank) extracted from the original 16kHz audio with a window size of 25ms and a 10ms shift.

\begin{figure}[htb]
  \centering
  \centerline{\includegraphics[width=5cm]{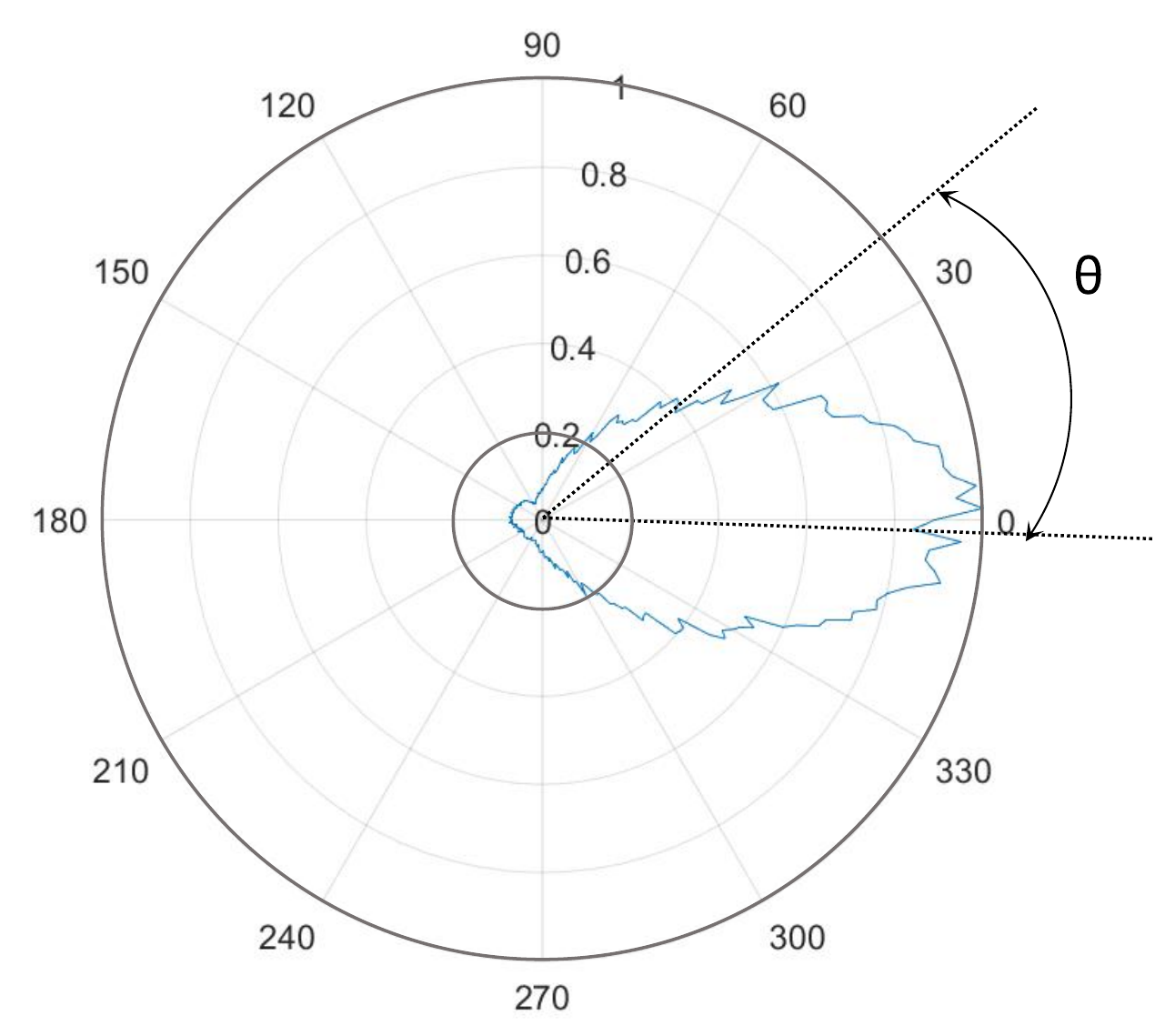}}
\caption{A diagram of circular array microphones.}
\label{fig:spatial_vector}
\end{figure}

\subsection{Spatial embedding}
The spatial embeddings are extracted by a DOA estimator. As shown in Figure~\ref{fig:spatial_vector}, the look-direction of each beamformer is uniformly distributed around a circle, so as to generate the spatial separation representations of the beamformer. Superdirective beamforming (SDB) \cite{PanChao804} algorithm is used to extract spatial representation named s-vectors on sliding windows. We set the window length and window shift to 1s and 0.5s respectively, which is consistent with speaker embedding extractor. The signal of the \textit{C}-channel microphone array is denoted as \textbf{x}$_{c}$(\textit{t})(\textit{c}=1,2,...,\textit{C}). The final output of the beamformer can be expressed as:
\begin{center}
\begin{equation}
\mathbf{Y} (\omega )=\sum_{c=1}^{C} \mathbf{h} _{c}^{*}(\omega ) \mathbf{x} _{c} (\omega )
\end{equation}
\end{center}
Eq.1 can be simplified into the following matrix form:
\begin{center}
\begin{equation}
\mathbf{Y} (\omega )=\mathbf{H}^{H}(\omega )  \mathbf{X} (\omega ) 
\end{equation}
\end{center}
where $ \mathbf{H}(\omega ) =\left [ \mathbf{h}_{1} (\omega ) ,\space \mathbf{h}_{2} (\omega ), \space ..., \space \mathbf{h}_{C}(\omega )\right ]^{T} $ is the beamforming filter of the array, $ \mathbf{X}(\omega ) =\left [ \mathbf{x}_{1}(\omega ), \space \mathbf{x}_{2}(\omega ), \space \cdots ..., \mathbf{x}_{C}(\omega )\right ]^{T} $ is the received signal, and superscript `\textit{H}' represents conjugate transpose. Given $ \mathbf{H}(\omega ) $, the response in space domain of the array is:
\begin{center}
\begin{equation}
\mathbf{B} (\omega,\mathbf{\theta})=\mathbf{H}^{H}(\omega)\mathbf{d} (\omega ,\mathbf{\theta })
\end{equation}
\end{center}
where $ \mathbf{\theta } $ indicates direction, $ \mathbf{d} (\omega ,\mathbf{\theta}) $is the steering vector of the array.
For suppressing the noise that is outside the steering direction, the idea of SDB is to maximize the directivity factor. Combined with no distortion constraint in the desired direction, SDB is transformed into an optimization problem:
\begin{center}
\begin{equation}
\min_{\mathbf{H}(\omega)} \mathbf{H} ^{H} (\omega)\mathbf{R_{NN}}\mathbf{H}  (\omega ) \quad s.t. \mathbf{H}^{H}(\omega) \mathbf{d} (\omega ,\mathbf{\theta _{0} } )=1
\end{equation}
\end{center}
where $\theta _{0}$ is look-direction of target speaker. The filter coefficient can be obtained as:

\begin{center}
\begin{equation}
\mathbf{H} (\omega)=\frac{\mathbf{{R_{NN}}^{-1}} (\omega )\mathbf{d}(\omega ,\mathbf{\theta _{0} } ) }{\mathbf{d^{H} }(\omega ,\mathbf{\theta _{0} } )\mathbf{{R_{NN}}^{-1} }(\omega )\mathbf{d}(w,\mathbf{\theta _{0} } ) }
\end{equation}
\end{center}
where $ \mathbf{R_{NN} }(\omega ) $  is the covariance matrix of noise.
The dimension of filter is (\textit{N}, \textit{C}, \textit{K}) in DOA estimator. In our experiment, according to the number of audio channels, we set \textit{C}=8. The order of FIR in the SDB was \textit{K}=128 in our experiment. We also set the number of FIR filter \textit{N} to 120 (a value optimized experimentally). These filters in DOA estimator were artificially designed as \textit{N} depth-wise convolution with 128 kernels size. As shown in Figure \ref{fig:system_overview}, when the 8-channel raw audio is fed to the estimator, single-channel signals from \textit{N} directions can be obtained. By normalizing the energy of signals in \textit{N} directions, we extract spatial embedding (s-vector) of segmented speech. The spatial embedding \textbf{e$_s$}=\{e$_{s1}$,e$_{s2}$,...,e$_{sN}$\} denotes the energy distribution of the different look-direction beamformer, where e$_{sn}$ represents the probability of speaker in \textit{n} direction. By increasing \textit{N}, higher resolution can be obtained, and we find that suitable value of \textit{N} will improve the robustness of s-vector.

\subsection{Later fusion}
In the embedding fusion module, we perform late fusion method \cite{kang2020multimodal} to construct a separate similarity matrix \textbf{A$_{sx}$}. We score the cosine similarity of embeddings in pairs to get similarity matrix. We then use the following formula to yield the fused similarity matrix which act as the input of clustering module:
\begin{center}
\begin{equation}
\label{eq:fusion}
{\mathbf{A}_{sx}}=a {\mathbf{A}_{x}} +(1-a) {\mathbf{A}_{s}}
\end{equation}
\end{center}
where \textbf{A$_{x}$} and \textbf{A}$_{s}$ denoted the cosine similarity matrix of x-vector and s-vector, respectively.

\subsection{Clustering}

In this stage, we perform normalized maximum eigengap spectral clustering (NME-SC) \cite{park2019auto} to obtain speaker labels of segments. NME-SC can estimate the number of clusters adaptively.

\begin{table}[]
\centering
\caption{The topology of DMSNet architecture.}
\begin{tabular}{cc}
\hline
Layer & Output Size \\ \hline
Input Layer & $L\times C$ \\
ASDB Block& $T\times D\times 1$ \\
Conformer & $T\times D\times 1$ \\
FC1 & $T\times 128$ \\
FC2 & $T\times 2$ \\
Softmax & $T\times 1$ \\ \hline
\end{tabular}
\vspace{-0.4cm}
\end{table}

\begin{figure}[!htb]
  \centering
  \centerline{\includegraphics[width=8.5cm, trim=200 0 400 0,clip]{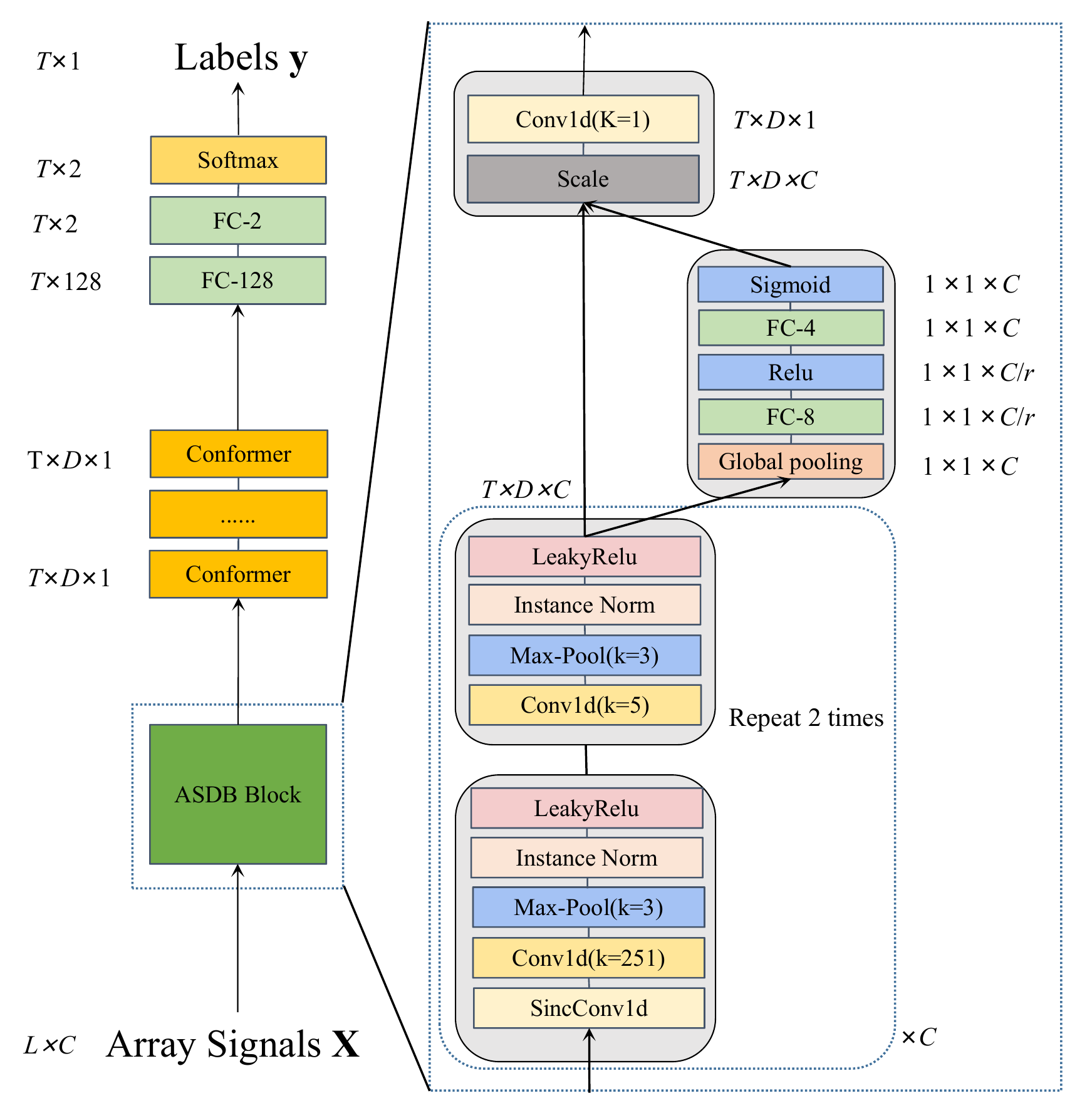}}

\caption{DMSNet is a multi-channel sequence-to-sequence architecture used for sequence labeling task. \textit{L} is the length of raw audio, \textit{C} is the number of channel and \textit{T} is the frame number of each chunked audio. The parameter \textit{r} is a reduction ration for controlling the computational cost of SE-Block. Instance Norm denote Instance Normalization layer.}
\vspace{-0.2cm}
\label{fig:DMSNet}
\vspace{-0.4cm}
\end{figure}

\subsection{Overlapped speech detection} \label{subsection:osd} 

The duration of overlapped speech accounts for a large proportion in AliMeeting corpus. In our system, we take two-stage OSD method to solve overlapped speech problem. The first stage is to detect overlapped region of speech. Then, in the second stage, we use the output of heuristic algorithm \cite{otterson2007efficient} to obtain secondary speaker labels. The secondary speaker labels of non-overlapped region will be removed. In the first stage, we propose a novel neural network architecture named DMSNet to detect overlapped speech of multi-channel audio. DMSNet is a multi-channel sequence-to-sequence architecture which can address the sequence labeling task which matches a raw audio \textbf{X} = \{\textbf{x}$_1$, \textbf{x}$_2$,..., \textbf{x}$_C$\} to the corresponding label sequence y where \textbf{x}$_c$ = \{x$_1$, x$_2$,...,x$_L$\} is of \textbf{L} length raw waveform and \textbf{y} = \{y$_1$, y$_2$,...,y$_T$\} is \textbf{T} frames label. The labeling principle is that y$_t$=0 if there is non-overlapped at time step \textit{t} and y$_t$=1 if there is overlapped. We use pyannote.audio toolkit \cite{bredin2020pyannote} to built this module. 

As shown in Figure~\ref{fig:DMSNet}, DMSNet consists of attention superdirective beamforming (ASDB) block and Conformer encoder. Inspired by SDB algorithm which designs different artificial filters for channels to enhance the signal, we propose a learnable ASDB block to overcome many shortcomings of the SDB. In ASDB block, the 8-channels raw audio is split into segments by sliding window and the feature of segments is extracted by SincNet \cite{ravanelli2018speaker}. The weights of each channels are self-adapted, which make the module learn the spatial information of speech better. Squeeze-and-excitation (SE) block \cite{hu2018squeeze} is applied to learn the weights of each channel and proposed to improve the representational power of a network enabling it to perform dynamic channel-wise feature. Then, we use 1-dimension convolution whose kernel size is 1×1 to combine the latent spatial feature of different channels. We also adopt 24 layers Conformer as encoder to capture speaker-spatial information of signal from the entire sequence. Finally, the classification layer including linear and softmax layers output the label. The complete architecture for DMSNet is shown in Figure~\ref{fig:DMSNet}. We train the architecture with binary cross entropy (BCE) loss.

\begin{table*}[]
\centering
\caption{Comparison of different OSD architecture. SDB-SincNet denotes that the inputs of SincNet were beamformed by SDB.}
\label{tab:OSD}
\begin{tabular}{cccccccc}
\hline
ID & Architecture & Extraction block & Encoder & DetER(\%) & Accuracy(\%) & Precision(\%) & Recall(\%) \\ \hline
M1 & PyanNet{ \cite{bredin2020pyannote}} & SDB-SincNet & Bi-LSTM & 42.57 & 91.61 & 85.23 & 70.06 \\
M2 & DMSNet-SDB(ours) & SDB-SincNet & Conformer & 35.48 & 92.97 & 87.22 & 75.58 \\
M3 & DMSNet-L(ours) & ASDB & Bi-LSTM & 36.68 & 92.69 & 88.18 & 73.12 \\
M4 & DMSNet(ours) & ASDB & Conformer & \textbf{32.47} & \textbf{93.53} & \textbf{89.22} & \textbf{76.81} \\ \hline
\end{tabular}
\vspace{-0.2cm}
\end{table*}
\section{Data preparation}
\label{sec:data_preparation}
The dataset was collected by 8-channel microphone array, of which 212 audio samples are training set (Train), 8 audio samples are evaluation set (Eval).  The number of speakers within one meeting session ranges from 2 to 4. Our use of training set in this challenge is as follows:
\begin{itemize}
\item Speaker embedding extractor: We take CN-celeb1 (793 speakers) and CN-Celeb2 (2000 speaker) \cite{fan2020cn} as the training set containing 2793 speakers in total (the CN-Celeb1-test part is excluded from training). 
\end{itemize}
\begin{itemize}
\item Overlapped speech detection: We use AliMeeting training set to train OSD models.
\end{itemize}

\section{Experimental result and discussion}
\label{sec:experimental_result}
\subsection{OSD methods}
The first experiment in this work is to explore the importance of different components in DMSNet. As shown in Table~\ref{tab:OSD}, we design different OSD architectures with different extract block and encoder layers. The inputs of models are chunk 2s audio. The 60-dimensional speech features were extracted by SincNet. We perform SDB to enhance the 8-channel audio on AliMeeting evaluation set using PyanNet architecture and DMSNet-SDB architecture. The encoder of DMSNet-L is 2 layer Bi-LSTM with 128-dimension hidden units. Considering that the filter and time delay of each channel is discriminative in SDB method, we make the weight of each channel different in ASDB. In DMSCNet-L architecture, we take ASDB as extraction block. DMSNet consists of an ASDB Extraction block and 24 layers Conformer encoder whose topological structure are the same in all OSDs.

We report the overall accuracy (overlapped and non-overlapped) precision, recall and DetER on the evaluation set. DMSNet achieve the best performance among these architectures and ASDB block is a necessary component of DMSNet. The purpose of FBS is to enhance the target speaker signal, while inhabits the rest of the speakers signals. As a result, the spatial information of array signals is lost. ASDB extraction block is based on channel attention to extract the spatial feature of speaker, which is suitable for OSD task. Conformer encoder is good at capturing local and global dependencies of sequence. By performing DMSNet, we achieve 93.53\% accuracy and 32.47\% DetER on evaluation set. 

\begin{table}[]
\centering
\caption{Experimental result for various diarization system with different speaker embeddings. The collar size of DER is 0.25s.}
\label{tab:embedding_fusion}
\begin{tabular}{ccc}
\hline
Embedding & \multicolumn{2}{c}{DER(\%)} \\ \cline{2-3} 
 & Score /w overlap & Score w/o overlap \\ \hline
s-vector & 17.13 & 5.14 \\
x-vector & 13.81 & 0.87 \\
sx-vector & \textbf{13.45} & \textbf{0.57} \\ \hline
\end{tabular}
\vspace{-0.5cm}
\end{table}

\subsection{Spatial-speaker embedding fusion}
The second experiment is designed to investigate different speaker embedding as input to the clustering module. In order to evaluate the robustness of various embeddings, we perform NME-SC for speaker diarization with different embeddings. In our diarization systems, the window length is 1s and the window shift is 0.5s. In this experiment, we use oracle VAD label and we evaluate the system with different scoring options. As shown in Table~\ref{tab:embedding_fusion}, the individual spatial embeddings are not robust for clustering. According to Eq.~\ref{eq:fusion}, we fused the spatial embeddings and speaker embedding. After tuning on training set, we set the \textit{a} to 0.95. By comparing with different scoring options in Table~\ref{tab:embedding_fusion}, we found that overlapped speech caused the high DER. When we ignored the overlapped speech, the fusion method resulted in 34\% relative improvement on evaluation set, which proved the robustness of spatial-speaker embedding.

\begin{table}[]
\centering
\caption{Detailed experimental results for various diarization system with different OSD module. The collar size of DER is 0.25s.}
\setlength{\tabcolsep}{1.2mm}{
\label{tab:diarization_systems}
\begin{tabular}{cccccc}
\hline
ID & OSD & MISS(\%) & FA(\%) & SpkErr(\%) & DER(\%) \\ \hline
S1 & - & 12.9 & 0 & 0.5 & 13.45 \\
S2 & Oracle & 1.8 & 0.0 & 4.3 & 6.10 \\ \hline
S3 & PyanNet & 5.5 & \textbf{0.3} & 2.9 & 8.75 \\
S4 & DMSNet-SDB & 3.6 & 1.0 & 3.6 & 8.17 \\
S5 & DMSNet-L & 4.2 & 1.1 & \textbf{2.8} & 8.13 \\
S6 & DMSNet & \textbf{3.6} & 0.6 & 3.5 & \textbf{7.64} \\ \hline
\end{tabular}}
\vspace{-0.55cm}
\end{table}

\subsection{Speaker diarization systems}
The third experiment is to apply the proposed DMSNet OSD for speaker diarization on AliMeeting evaluation set. To investigate the importance of OSD, we also evaluate a speaker diarization system with oracle OSD label. For System 1$\sim$6, we use spatial-speaker similarity matrix as input to clustering module. Table~\ref{tab:diarization_systems} shows the results of experimental results of our systems with different OSD modules. System 1 uses NME-SC to assign the most probability speaker label to segment. The result of System 2 proves the importance of OSD module, i.e. there is still a room for improvement on the detection of overlap. Compared with System 3, the result of System 6 suggests that using DMSNet can significantly improve the performance on DER.

\section{Conclusions}
\label{sec:conclusions}
In this paper, we propose a spatial-aware speaker diarization system for multi-channel multi-party recording in real-world meeting. We present two methods to improve the performance of speaker diarization system for multi-channel audio. We also propose a sequence-to-sequence architecture named DMSNet for OSD. DMSNet is a channel-wise neural network architecture which adaptively extracts latent spatial information of array audios by modeling interdependencies between channels. We also perform a DOA estimator to extract spatial embedding of speaker. Experimental results show that DMSNet outperforms the existing overlapped speech detection method, and the results demonstrate its effectiveness in speaker diarization systems.  

\section{Acknowledgements}
Thanks to the National Natural Science Foundation of China (Grant No.61876160 and No.62001405) for funding.

\bibliographystyle{IEEEtran}

\bibliography{mybib}


\end{document}